# Is Research Funding Always Beneficial? A Cross-Disciplinary Analysis of UK Research 2014-20

Mike Thelwall
Statistical Cybermetrics and Research Evaluation Group, University of Wolverhampton, UK.
https://orcid.org/0000-0001-6065-205X m.thelwall@wlv.ac.uk

Kayvan Kousha
Statistical Cybermetrics and Research Evaluation Group, University of Wolverhampton, UK.
https://orcid.org/0000-0003-4827-971X k.kousha@wlv.ac.uk

Mahshid Abdoli
Statistical Cybermetrics and Research Evaluation Group, University of Wolverhampton, UK.
https://orcid.org/0000-0001-9251-5391 m.abdoli@wlv.ac.uk

Emma Stuart
Statistical Cybermetrics and Research Evaluation Group, University of Wolverhampton, UK.
https://orcid.org/0000-0003-4807-7659 emma.stuart@wlv.ac.uk

Meiko Makita
Statistical Cybermetrics and Research Evaluation Group, University of Wolverhampton, UK.
https://orcid.org/0000-0002-2284-0161 meikomakita@wlv.ac.uk

Cristina I Font-Julián
Department of Audiovisual Communication, Documentation and History of Art, Universitat Politècnica de València, Spain
https://orcid.org/0000-0003-2351-4816 crifonju@upv.es

Paul Wilson
Statistical Cybermetrics and Research Evaluation Group, University of Wolverhampton, UK.
https://orcid.org/0000-0002-1265-543X pauljwilson@wlv.ac.uk

Jonathan Levitt
Statistical Cybermetrics and Research Evaluation Group, University of Wolverhampton, UK.
https://orcid.org/0000-0002-4386-3813 j.m.levitt@wlv.ac.uk

The search for and management of external funding now occupies much valuable researcher time. Whilst funding is essential for some types of research and beneficial for others, it may also constrain academic choice and creativity. Thus, it is important to assess whether it is ever detrimental or unnecessary. Here we investigate whether funded research tends to be higher quality in all fields and for all major research funders. Based on peer review quality scores for 113,877 articles from all fields in the UK's Research Excellence Framework (REF) 2021, we estimate that there are substantial disciplinary differences in the proportion of funded journal articles, from Theology and Religious Studies (16%+) to Biological Sciences (91%+). The results suggest that funded research is likely to be higher quality overall, for all the largest research funders, and for all fields, even after factoring out research team size. There are differences between funders in the average quality of the research they support, however. Funding seems particularly beneficial in health-related fields. The results do not show cause and effect and do not take into account the amount of funding received but are consistent with funding either improving research quality or being won by high quality researchers or projects. In summary, there are no broad fields of research in which funding is irrelevant, so no fields can afford to ignore it. The results also show that citations are not effective proxies for research quality in the arts and humanities and most social sciences for evaluating research funding.
**Keywords**: Research funding; research grants; research quality; academic careers; Scientometrics

# 1 Introduction

The amount of time spent writing and managing grants occupies a substantial amount of academic time, but it is not clear that the benefits outweigh the costs in all fields. In some cases, researchers may be unable to experiment without funding, but scholars not needing new equipment, resources or time buyout may be able to work equally well without it. Nevertheless, little is known about the proportion of academic time spent on grant writing or other specific tasks, so it is difficult to weigh the benefits of funding against its costs. Whilst many studies report the proportion of time spent by academics on research, teaching, and administration (Bentley & Kyvik, 2012), they rarely ask about grant writing as a separate activity. One exception estimated that each Australian National Health and Medical Research Council grant proposal took 38 working days (nearly 2 months), or 28 for a resubmitted proposal. In 79% of cases, this effort was unrewarded (Herbert et al., 2013), so Australian researchers collectively spent between half a year and a full year writing grant proposals for each one funded. A survey of 12 European countries found that between 51% and 84% of academics (71% in the UK) wrote grant proposals or otherwise responded to calls for proposals each year (Drennan et al., 2013). This work is highly stressful, affecting work-life balance in almost all applicants, but deemed necessary partly due to institutional pressure and expectations from colleagues (Herbert et al., 2014). There have also been claims that the constant pressure to win grants undermines the quality of research, particularly in situations where ongoing employment is funding-dependant (Fumasoli et al., 2015). Thus, it is important to assess whether the benefits of funding always outweigh the costs. This article focuses on the narrower issue of whether funding is always beneficial, in the sense of being associated with higher research quality outputs.

Whilst the time and equipment for early scientific research was self-financed or informally supported by benefactors, the system of competitively awarding national grants for future research emerged from the prize system (for previous achievements) in France before the first world war (Crosland & Galvez, 1989). Over the last half century, university funding in many countries has changed from being awarded unconditionally for the benefit of science, albeit with a focus on government priority areas, such as defence, to being mainly accountable and harnessed for societal benefits (Banchoff, 2002; Demeritt, 2000; Mustar & Larédo, 2002; Lepori et al., 2007), such as medical priority areas (Xu et al., 2014). Resource seeking behaviours ("academic capitalism": Slaughter & Leslie, 2001) are long established as the norm in several major research countries (Johnson & Hirt, 2011; Metcalfe, 2010). Research funding is now primarily awarded for achievements (i.e., performance-based funding: Hicks, 2012) or future promise, through competitive grants (OECD, 2014). This is supplemented by incentives to seek finances from industry and other non-academic sources to research for non-academic benefits (Laudel, 2005). External funding is considered valuable despite the huge amount of lost time spent by experts writing grant proposals (Polster, 2007) and the potential to skew science (Thyer, 2011). National grant awards may aim to generally support promising research or researchers, or support research with societal benefits (e.g., Takemura, 2021).

The effect of funding seems likely to depend on the researcher, with field-related variations. For example, in specialities needing funding for any kind of research (e.g., areas of medicine or genetics), such as to employ enough assistants or to access equipment or consumables, a researcher without funding cannot conduct any research, so what do they do instead? If their role is not changed to teaching only (Edwards, 2022; Nyamapfene, 2018) or professional (e.g., clinical doctor), they might use any research time allowance to write grant proposals, accept consultancy or advisory roles, read new research, develop aspects of their skills, develop their research methods or theory, or devote more time to teaching, administration or other roles. They may also support others' funded research projects in a minor role. In some cases, they may also write short papers about aspects of their research process, such as ethics, or minor methods details. In contrast, other researchers may easily be able to conduct unfunded research, even though funding might improve their research with better equipment or larger teams. From a UK Research Excellence Framework (REF) perspective, funding might improve research productivity by supporting larger teams (allowing divisions of labour) and give the funded researcher a larger pool of publications from which to choose their best for REF evaluations. Thus, any comparison between funded and unfunded research pre-supposes that it is possible to research in a field without funding and necessarily excludes researchers that can only conduct funded research but did not receive funding in the period examined. The situation is complicated in the UK because university teaching budgets subsidise 13% of research (Olive, 2017) in a way that may not be recorded. Researchers with the choice may prefer unfunded research because it gives them autonomy from funder goals and requirements (Edwards, 2022), particularly in the era of challenge led research (Olive, 2017).

Few studies have explicitly examined the relationship between research quality and funding, except with bibliometric proxies for quality, such as citation rates. One exception found that industry funded submissions to a 2006 Scoliosis conference not to have statistically significantly different peer review quality scores (Roach et al., 2008). Another showed that urogynecology randomised controlled trials had higher quality methods when they were funded (Kim et al., 2018). Although, as reviewed below, many studies have shown in different contexts that funded research tends to be more cited than unfunded research, it is not known whether this is true in all fields and if funded research also tends to be higher quality. Thus, the research questions below mostly address issues that have not previously been directly investigated. A question about citation counts is included to assess their value as a tool because of their use in most previous investigations of research funding.

- RQ1: How prevalent is research funding for UK journal articles and are there disciplinary differences in the answer?
- RQ2: Is funded research (UK journal articles) higher quality for all major research funders?
- RQ3: Do research funders support different quality research (UK journal articles)?
- RQ4: Is funded research higher quality irrespective of team size?
- RQ5: Are average citation counts effective proxies for average quality for externally funded research?

# 2 Background: Research funding types and benefits

Although there is little research about the relationship between funding and research quality, this relationship is affected by multiple: variations in funding types, the effectiveness of proposal reviewing, the impact of individual grants and their career benefits, and unfunded research. Whilst the focus here is on disciplinary differences and there are theories of the organisation of science that include the importance of funding (e.g., Whitley, 2000), none yet shed light on the relationship between funding and research quality. A partial exception is that grant review outcomes might be expected to be more unpredictable in fields with low level of agreement on the objects and methods of research (task uncertainty: Whitley, 2000), probably including most of the social sciences and humanities.

## 2.1 Types of funding

There are many types of research funding, in addition to recurring block grants. Each has its own unique characteristics, including the following.
- Goal specificity: Specific (e.g., test the efficacy of a given vaccine), loose (e.g., any data science research into modern slavery; social science research into economic migration), or open (any social science research).
- Competitivity: highly competitive (e.g., <1% success rate) to non-competitive (e.g., an industry contract built from an existing working relationship).
- Project-based or person-based (e.g., PhD student funding).
- Size (e.g., $12.8 billion for an international fusion experiment or £500 for travel).
- Qualifying applicant criteria (e.g., UK university, anyone).
- Rationale/source: Academic (develop knowledge) or non-academic (e.g., solve an industrial, health, or governmental problem).
- Length of funding: short-term (e.g., a small grant for a single quick study), medium term (e.g., a three-year project) or long term (e.g., a ten-year project, a lifelong research chair, a large facility designed to operate for half a century).
- Cost scope: solely for facilities, equipment, consumables, researcher time, overheads, or a combination of these, or free to be spent at the discretion of the awardee.
- Post-funding evaluation: needed or not.

These factors may affect whether a grant is beneficial to a researcher's output, so it is unsurprising that the performance of funded researchers varies between funding schemes, even for public funders (Wang et al., 2020). It is impossible to differentiate between funding types in practice for any comprehensive investigation into the impact of all types of funding on research. Nevertheless, relevant evidence has emerged from prior studies for two dimensions, as summarised below.

**Size**: The average size of individual grants has increased over recent decades, for example with long-term funding for large centres of excellence at the expense of sets of individual grants (Bloch & Sørensen, 2015; OECD, 2014). In the USA, block funded National Science Foundation centres do not seem to improve the journal outputs of members, although they do improve commercial partnerships (Gaughan & Bozeman, 2002). Smaller grants seem to help productivity more than larger grants for research centres (Bloch et al., 2016). Larger grants may direct the entire programme of a research group, opportunistically shifting their goals to align with potential sources of money (Jeon, 2019). There are many specialist sources of small grants. For example, a UK fisheries charity allocated £560,000 over 25 years in grants of up to £6,000 (Lyndon, 2018). Small grants seem likely to have different effects from medium and large grants, presumably through supporting individual small-scale studies. Smaller research awards lead to more citations overall for biological science research (Gallo et al., 2014).

**Rationale/source**: The source of funding received by a research group influences their research agenda (Tellmann, 2022). For example, German science and engineering research groups in 2000 getting a large share of their funding from large businesses, seed their research with ideas that may commercially benefit their funders. Similarly, development funding often aligns with political priorities, aims to generate self-sufficiency and nurtures local talent (Currie-Alder, 2015). In contrast, ideas for research groups mainly attracting public funding tend to come from within science (Hottenrott & Lawson, 2014). Having commercial funding at the same time as a non-commercial grant in the UK seems to suppress the benefits of the non-commercial grant, with reduced publication outcome increases (Hottenrott & Lawson, 2017). This illustrates the complexity of the relationship between funders and outcomes. US engineers with ongoing funding from mission-oriented government funding agencies (e.g., NASA) also produce fewer academic outputs (Goldfarb, 2008). In terms of quality, spinal research harnessed weaker types of evidence (e.g., case series) when it had industry funding but was more likely to report positive outcomes (Amiri et al., 2014).

## 2.2 Effectiveness of grant proposal peer review and characteristics of recipients

The peer review process for academic grants has received considerable scrutiny, often in terms of the characteristics of the awardees. This is relevant to the relationship between funding and research outputs because effective funding selection criteria may lead to better researchers being funded or receiving more funding. Although some evaluation of peer review grant selection focus on the achievements of the applicants, most concentrate on the proposal, checking its rationale, evaluating its validity and (often) match with funding criteria (Chubin, 1994; Franssen et al., 2018). For example, in The Netherlands, the decision process for three types of career grants is only slightly influenced by the scientific impacts of applicants and the opinions of external referees, with the promise of proposals and researchers being most important, as judged partly through interviews (Van Arensbergen & Van Den Besselaar, 2012). In contrast, a study of four types of grants awarded in Germany found that applicants' productivity associated with increased success rates for research foundations and industrial funding, but not for EU or government funding. The results also showed that research groups often focused on types of grants for which they had successes (Grimpe, 2012).

There is limited overall evidence of the effectiveness of peer review (Liaw et al., 2017), based on evaluations typically using citation indicators as a proxy for research quality or achievements. Nevertheless, an analysis of early career grant decisions for a biomedical research funder in Germany found that awarded applicants were more cited than rejected applicants and both were more cited than non-applicants (Bornmann & Daniel, 2006). An analysis of awarded biological grants found that higher scoring projects tended to produce more cited research (Gallo et al., 2014). Higher peer review scores for basic Hungarian research grants (Győrffy et al., 2020) and NIH grants (Fang et al, 2016) associated with slightly more outputs, although peer review scores do not reflect post-review bid improvements (Lindner & Nakamura, 2015). In contrast, for economic and social sciences research council grants in The Netherlands, whilst weak researchers tended to be rejected, awardees performed substantially worse in bibliometric terms than rejected researchers with similar scores (Van den Besselaar et al., 2009). This suggests that the research council process selected above average researchers but not the very highest performing, or that the funding was detrimental.

Many studies have found disparities in review outcomes that are suggestive of bias, whether deliberate or accidental, or systemic effects. Male researchers are more likely to be awarded grants than female researchers, sometimes because more apply, but often because of apparent biases in reviewing (Cruz-Castro et al., 2022; Tricco et al., 2017), such as by failing to consider career gaps for carer responsibilities. For US biomedical funding, the average age of grant winners has increased steadily over time, suggesting that young researchers may struggle to get their own ideas funded (Levitt & Levitt, 2017). Ethnic minority researchers are also less likely to win some types of funding (Cruz-Castro et al., 2022; Hayden, 2015). From a topic perspective, interdisciplinary research is less likely to be funded under most reviewing models (Seeber et al., 2022). In the USA, faculty are more likely to receive more and larger national research grants if they are more productive, at a private university, and within an Association of American Universities (i.e., research intensive) institution (Ali et al., 2010). For European grants, previously successful applicants at prestigious institutions are most likely to be awarded new funding (Enger & Castellacci, 2016). Funding tends to benefit research intensive universities in some countries (Horta et al., 2008; Jappe Heinze, 2023). In Quebec, Canada, a minority of researchers monopolise research funding, with diminishing returns compared to a more equal sharing of funding (Mongeon et al., 2016). All biases seem likely to reduce the effectiveness of the grant allocation process and hence, presumably, the overall benefits of funding.

## 2.3 *The impact of grants on research productivity and impact*

Funding could be expected to increase the productivity or impact of the funded researchers. The benefits of research funding are impossible to fully quantify, and it is difficult to generate meaningful statistics because of the lack of effective control groups in most cases, and particularly the ability of unfunded groups to receive funding from sources other than the one examined (Neufeld, 2016; Schneider & van Leeuwen, 2014). Most previous studies have analysed specific funding sources and assumed that the papers acknowledging them were primarily caused by the funding, whereas journal articles often draw upon a range of different long-term and short-term funding for equipment and different team members as well as specific project-based grants, at least for biomedical research (Rigby, 2011). Moreover, many studies do not distinguish between selection effects and funding effects (Neufeld, 2016): are funded researchers more productive because of the money or because better researchers/proposals were selected, or both? Moreover, all studies so far have had limited scope, there are different types of funding and disciplinary differences in funding uses and procedures so there is unlikely to be a simple relationship between funding and impacts. For example, larger funded studies may find it easier to get ethical approval to research in clinical settings (Jonker et al., 2011), reducing the number of unfunded studies.

Funding usually associates with increased research productivity, as measured by journal articles (Saygitov, 2018), although it is often difficult to differentiate between the results of funding or the success of the selection process (discussed above). In Japan, grants addressing government strategic priorities are successful at increasing research productivity, even after the funding period has ended (Shimada et al., 2017). In Luxembourg, national research grants increase productivity, but the effect disappears after five years (Hussinger & Carvalho, 2022). Argentinian research grants increase the number of articles (and average journal impact factors) of awardees compared to high scoring unsuccessful applicants

(Chudnovskyet al., 2008). For Canadian universities, 85% of science and engineering journal articles 1990-99 were funded, and the receipt of funding improved productivity (Godin, 2003). European Union Framework Programme funding also increases research productivity and funded collaborations increase productivity after the funding has finished (Defazio et al., 2009). For Canadian natural science and engineering research 1996-2010, funding improved productivity (and citation rates), even after accounting for collaboration (Ebadi & Schiffauerova, 2016). For small collaborative education grants for central USA medical colleges, awarded academics produced more outputs and collaborated more than unsuccessful applicants (El-Sawi et al., 2009). A systematic attempt to track down all funding sources for research from one university suggested that funding increased productivity but not citation impact, although it would be difficult to disentangle disciplinary differences in funding value with this data (Sandström, 2009). In contrast, commercial funding can slow academic publishing because of the need to write patents or produce other outcomes (Hottenrott & Thorwarth, 2011).

Funding usually associates with higher citation impact (e.g., Berman et al., 1995). Biomedical research with more funding sources tends to be slightly more cited, although this could be due to large team effects rather than multiple rounds of peer review or a greater total amount of funding (Lewison & Dawson, 1998; Rigby, 2011). Researchers funded by the Swiss research funding agency attract more citations per paper and write more papers (Heyard & Hottenrott, 2021). Research funded by Wellcome, the National Institutes of Health and the Medical Research Council is cited above average for its field and year (Thelwall et al., 2016). For Spanish virology, cardiology and cardiovascular researchers, funded research tends to be more cited (Álvarez-Bornstein et al., 2019). Early economics research (1978 and 1979) was more cited when funded (Peritz, 1990). Articles published between 2010 and 2016 in Astrophysics, Computer Science, Engineering, Environmental Studies, Mathematics, Medicine, and Nanotechnology journals tended to be more cited when funded (Yan et al., 2018). Articles and reviews in economics, computer science and medicine from 2015 in the Web of Science tended to be more cited when funded (Roshani et al., 2021). Scientists funded for basic research in New Zealand by the main academic source tend to have higher productivity and average citation impact compared to unfunded applicants (Gush et al., 2018). Articles from 2008 funded by the National Science Foundation are more cited than average for their field (Levitt, 2011).

There can be disciplinary differences in the relationship between funding and citation impact. For 80,300 Scopus-indexed Iranian publications from 2000-2009, only 13% declared funding and funded research was more cited in 20 out of 22 fields. The (surprising) exceptions were Biology/Biochemistry and Environment/Ecology (Jowkar et al., 2011). An analysis of 280 applicants for German early biomedical career funding found that grant winners disproportionately improved their citation rates in biology but not medicine (Neufeld, 2016). In Italy, the situation is complex. For the natural sciences, agricultural sciences and engineering, small amounts of funding from public and private research contracts and consultancies associates with higher citation impact research, but too much funding reduces citation impact, perhaps due to a loss of focus on academic goals. The situation is reversed for medicine, where a small amount of funding from public and private research contracts and consultancies reduces research impact but a large amount increases it (Muscio et al., 2017). The latter may reflect the large-scale funding needed for effective medical studies in many cases, with underfunded research also being underpowered.

Unfunded research can be highly cited, perhaps because it has more scope to be innovative, at least in fields like library and information science not needing expensive resources (Zhao, 2010). Grants may constrain academic freedom, which is a particular threat to the role of social science research in challenging authority and in being free to interpret results free from external pressures (Kayrooz et al., 2007). In fields where the data or source material are freely available to many researchers, such as scientometrics, history and theology, funding may limit research by imposing a goal, such as an evaluation of a specific region, country, or organisation in scientometrics. This may direct research towards more applied and routine goals rather than more far-reaching basic research goals. It is therefore unsurprising that citation impact does not always increase with funding (Sandström, 2009), as in the case of radiology research in 2016 (Alkhawtani et al., 2020). Similarly, with open calls in Denmark and Norway, successful applicants do not increase their average citation impact but do increase productivity (Langfeldt et al., 2015).

## *2.4 Career benefits of research grants*

Some grants target researchers rather than projects, aiming to help them begin a successful research career or transition an existing career to a higher level. There is bibliometric evidence that some of these are successful in this regard. The Cheung Kong Scholars Award in China, for example, aims to support the country's best scholars in all fields. Receipt of this award associates with an increased publication rate, more citations, increased collaboration, and more leadership roles in research projects (Liu et al., 2018). Recipients of Danish research council grants are more productive during the funding period, had career-long enhanced status, and were twice as likely to become full professors than rejected applicants (Bloch et al., 2014). In line with this, European Research Council (ERC) Starting Grants have been described as career defining status bestowing events (Edlund & Lammi, 2022). Just over half of all National Institute of Health grant recipients became consistently successful at attracting new grants, building a well-funded career, with high scores in their first application being a good predictor of this (Haggerty & Fenton, 2018).

Awardees of a US scheme supporting postdoctoral fellowships abroad were more successful at generating international collaborations than a comparator group and often benefitted from long term relationships with their collaborating partners (Martinez et al., 2016).

## *2.5 Levels and types of unfunded research*

Most research in the previous century was unfunded, at least as reported in journals. An early study of 900 journal articles in three medical journals from 1987, 1989 and 1991 found high levels of unfunded research (at least without declared funding sources): internal medicine (60%), pathology (62%), and surgery (74%) (Berman et al., 1995). Similarly, in 1987, 1989, and 1991, 84% of journal articles by pathologists were unfunded (Borkowski et al., 1992) and 63% of emergency medicine articles were unfunded in 1994 (Ernst et al., 1997). In 1992, however, only 23% of internal medicine and neurology journal articles were unfunded (Stein et al., 1993).

Early unfunded research was often different from funded research. Three quarters of unfunded internal medicine and neurology journal articles were based on routine patient care, but the remainder incurred costs that could not be accounted for by named funding sources (Stein et al., 1993). Within paediatric neurology, busy senior staff enlist students to carry out investigations into "clinically oriented questions of limited scope" to the benefit of both and requiring little or no funding (Bodensteiner, 1995), which may explain the origins of early unfunded medical-related articles. US and Canadian psychiatric research from 1992 without external funding sometimes had minor internal funding sources, tended to occur in the most productive institutions, leveraged easily available patient populations, and focused on two main topics (Silberman & Snyderman, 1997). An early study found no evidence that funding increased the prevalence of quantitative methods in sociological research, however (Platt, 1996).

Research funding may have become more widespread this century, at least in medicine, with more funding and more recording of funding. A survey of authors publishing in the Journal of the American College of Cardiology between 2007 and 2009 found that partially funded research was the norm (44%) compared to fully funded (26%) and unfunded (30%) (Mai et al., 2013). Nevertheless, some specialities lack funding: 88% of research on clinically relevant digital measures 2019-2021 (Shandhi et al., 2021) and 73% of radiology articles 2001-2010 were unfunded (Lim et al., 2012). This suggests funding differences between these specialties, although 84% of the radiology papers published were clinical (Lim et al., 2012) so the difference may be partly due to easier publishing for clinical research.

A few studies have compared funded with unfunded research types this century. For Spanish virology, cardiology and cardiovascular researchers, unfunded research was hospital-based and clinical, suggesting that the study has been internally supported by hospital resources (Álvarez-Bornstein et al., 2019). Unfunded investigations may tend to be desk research or other cheaper types, including secondary data analysis (Vaduganathan et al., 2018), guidelines (Goddard et al., 2011), review articles (e.g., Imran et al., 2020), retrospective records-based analyses (e.g., Brookes et al., 2021; Sedney et al., 2016), small case studies (e.g., Qi & Wei, 2021), or analytical/theoretical/opinion papers without primary data (Underhill et al., 2020). In nursing, evidence-based practice research may often be unfunded because the data analysed may come mainly from investigators' daily work roles (Higgins et al., 2019). Researching may be a compulsory part of some higher-level courses, such as for radiology, and this may result in many small-scale unfunded studies by educators and learners (Johnson et al., 2002). In medicine, unfunded research may be disproportionately from general practitioners compared to hospital doctors because they lack the infrastructure to obtain and maintain large grants (van Driel et al., 2017). Unfunded American College of Cardiology authors were typically younger (under 40), not from the USA, and with a clinical rather than a basic research specialty (Mai et al., 2013).

Although, as reviewed above, funded research tends to be more cited in most contexts, unfunded research can still be of the highest quality: 30% of key papers for physics, chemistry, and medicine Nobel Prize winners 2000-2008 declared no funding (Tatsioni et al., 2010). Similarly, 89% of the most cited rhinoplasty articles published by 2015 were unfunded (Sinha et al., 2016). Unfunded trials tended to have larger effect sizes than funded trials for lower back pain, but this raised the possibility that the unfunded trials were deficient in quality control (Froud et al., 2015). This partly aligns with unfunded nutrition and obesity peer-reviewed journal articles 2001-2011 being more likely to overstate their findings in their abstracts (Menachemi et al., 2013).

# 3   Methods

## *3.1   Data*

For this study, UKRI gave us the preliminary scores from March 2022 of all 148,977 journal articles submitted to REF2021, excluding those from the University of Wolverhampton. Each article had been given a quality score by at least two out of over 1000 expert assessors, with the grades being 1* (nationally recognised), 2* (internationally recognised), 3* (internationally excellent), and 4* (world leading). The grades reflect originality, significance, and rigour, with different guidelines for these from each of four overseeing Main Panels. There was norm referencing between assessors within each of the 34 field-based Units of Assessment (UoAs). The scores are used to direct all UK university block grant funding and the peer review process is carefully managed, although the reviewers are not experts in all areas that they need to assess.

The REF articles were matched against Scopus records by DOI comparisons (n=133,218). REF articles without a DOI in Scopus were matched instead by title, with a manual check to accept or reject all potential matches (n=997). The Scopus record was used for funding and citation information. Scopus cross-references information in articles with funding information gained from other sources populating its funding database (McCullough, 2021). Scopus reports a single funder for each paper, at least through its Applications Programming Interface (API), as used to gather the data, although some studies have multiple funders. Thus, the funder-level results reported here are based on incomplete data.

Some of the articles were given multiple grades from the same or different UoAs. This is possible because each author is entitled to submit between 1 and 5 outputs for which they are a co-author. For analysis, duplicate articles were removed within the grouping analysed (UoA, Main Panel, or all). When an article had received different scores, it was given the median or a random median when there were two.

Bibliometric databases record funding information extracted from articles (Rigby, (2011) and Scopus was used for this information because of its wider coverage than the Web of Science (Martín-Martín et al., 2021) and because Google Scholar does not extract funding information. Scopus started systematically indexing funding in 2011 (Rigby, 2011) so its data should be mature for the REF period 2014-20. Funding information in academic articles might be in a separate "Funding sources" section, in the acknowledgements, or as a footnote alongside author information. The acknowledgement section was a common place for funding information (Paul-Hus et al., 2017), before the rise of the dedicated funding section.

Some article funders were universities, suggesting that the authors had been allocated internal university money for their research or that it was unfunded but recorded as university-funded to reflect employers allowing research time for the scholars involved, or for university policy reasons. Since it was not possible to distinguish between the two, for the regression analysis, research was classed as unfunded if the funder was a university, irrespective of country. For this, we checked the 4042 funders and classified 1317 of them as internal university or research institute funding (e.g., Weizmann Institute of Science) and 2725 as external funding (e.g., American Mathematical Society). After this stage, research was classified as externally funded if it declared a funder in Scopus and the funder name was not one of the 1317 universities found. When funding information was present (e.g., a grant number) but no name for the funder was given, it was assumed to be externally funded.

## 3.2 Data quality checks

To check whether the Scopus API funding information was accurate, for six UoAs chosen to represent different field types we read the article for details of support for the research. For each UoA, 100 unfunded articles, 100 university-funded articles, and 100 non-university funded articles were selected with a random number generator for checking (or 100%, when fewer than 100). The checking was performed by two people, the first author for all and either ES, MM, or MA. Publisher versions of articles were checked for funding information except when the preprint was online with funding information. In one case (Theology, unfunded) we were unable to obtain the article through any method (including inter-library loans) and it was substituted with the next article selected by the random number generator. A study was counted as university funded if the only funding source mentioned was university-based. It was recorded as externally funded if any non-university funding source was mentioned.

Funding could be mentioned in multiple places, although a "Funding" section or an "Acknowledgements" section at the end of the article were common, and a "Disclosure of Interests" end section sometimes also included funding information. Other places included first page footnotes, last page footnotes (rare), a notes section at the end of the article, the first paragraph of the article (rare), and the last paragraph of the conclusions (common in Physics, one example in Theology). Articles sometimes declared that the research was unfunded, usually within a funding section, and sometimes in a disclosure of interests section. In one case, a funding section declared that the research was unfunded but the acknowledgements section thanked a funder, so this was coded as funded. Some articles included author biographies that might have mentioned funding sources but never did.

Funding statements varied in length from short declarations of the funder name and grant number to several paragraphs of thanks. In some fields it was common to thank departments hosting a visit or seminar and current and former employers. An article was classed as funded if this was stated directly (e.g., "funded by") or if it was suggested by the context, such as by naming a research funding organisation or thanking one for an unspecified type of "support". Acknowledgements of support from universities were not counted as funding if these seemed to be minor and routine, such as hosting a visit or supporting a seminar. University support was counted as funding if the term "funding" was mentioned or "grant" or it was obvious from the context that a financial transaction had occurred, as in the case of a PhD studentship. In a few cases, support in kind was provided, such as through access to equipment, but this was not counted as funding. Research described as part funded was recorded as funded.

Whilst in some cases the article appeared to be the primary outcome of a grant, in most cases the relationship between the funding and the output was less clear. For example, the article could be one of the outputs of a PhD studentship or Leverhulme Trust fellowship. Many articles had authors with differing funding sources, suggesting that the study itself had not been funded but had been made possible by funding given to the participants. Such studies were counted as funded. In Medicine and Physics, for example, long paragraphs often recorded the financial support given to all participants as well as the equipment used and the study itself.

The information found manually is unreliable. A funded article may have no declaration within the text if the author forgot or the journal style or field norms discourage it. Checks were made of cases where Scopus recorded a funder but the article didn't mention one. These checks found examples where Scopus was correct because the article was listed on a funding website as an output of the grant. Although Scopus has reported that its funding information is imported from the acknowledgement sections of articles (Beatty, 2017), it seems likely that it now automatically links articles to funding records from elsewhere and might also perform wider searches of article text. In other cases the funding was plausible because the scholar thanked the same funder on a different output at a similar time or listed the funder on their online CV. Scopus also seemed to have listed incorrect funders in at least two cases: the wrong funder altogether in one case, and a university in another case where the author had included an acknowledgement that an earlier version had been presented at a seminar at that university.

Comparing the Scopus API information with manual checks, the Scopus API results were always imperfect and substantially misleading in some cases (Figure 1). Almost all Clinical Medicine and Physics articles were externally funded (i.e., at least one non-university funder) even if the Scopus API listed none. In these cases, Scopus had presumably not found where the funding was listed in the article. Physics article funding statements were often in the last paragraph of the conclusions, where they may have been missed. For all six fields, most articles classed as university funded (i.e., the Scopus API funder was apparently a university) were externally funded. This typically occurred because the Scopus API reports only one external funder, and the manual checks classed an article as externally funded if any of the funders were not universities. In four cases, most Scopus API results were correct for unfunded and externally funded articles, however. This information should be taken into consideration when interpreting the results.

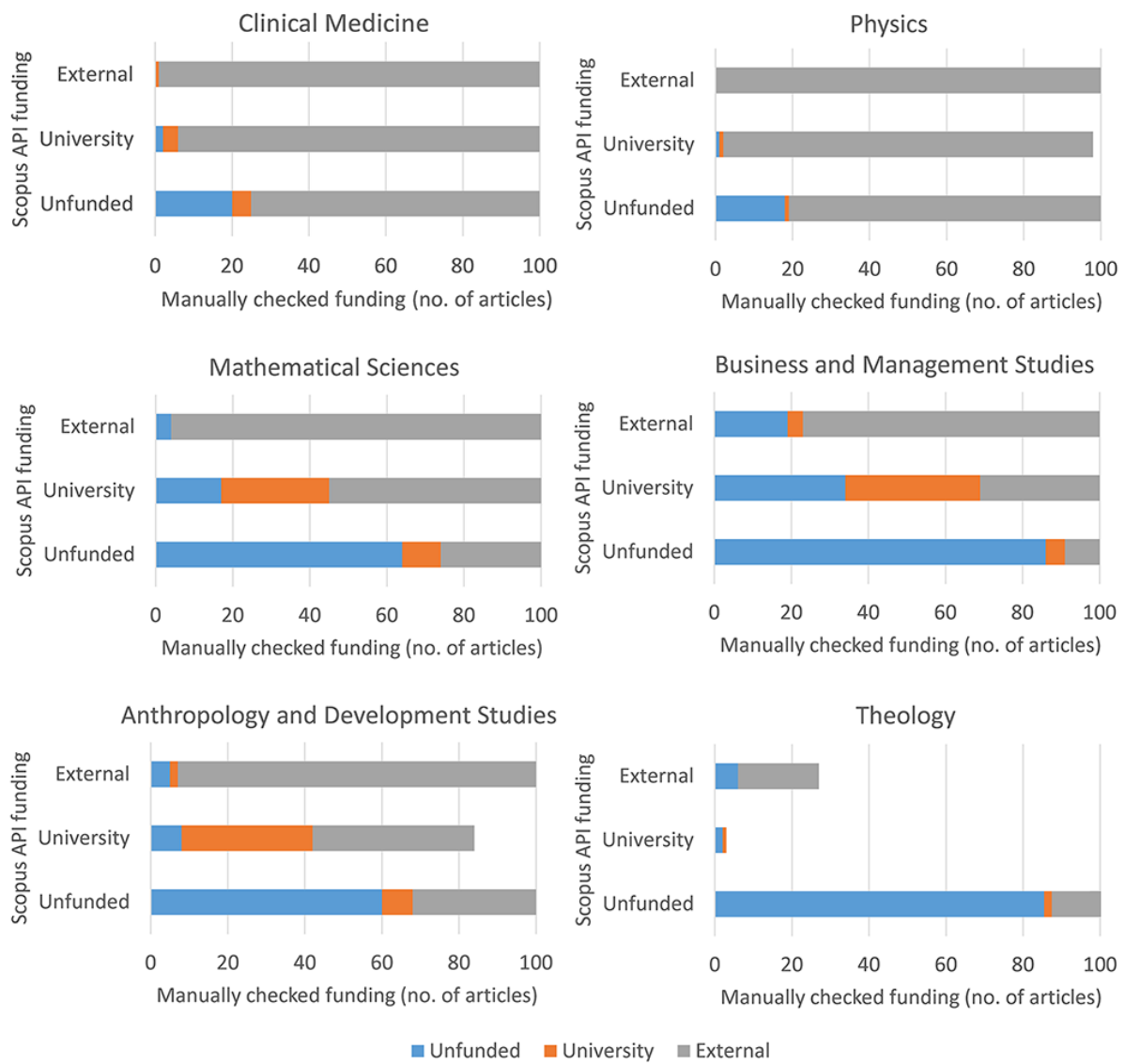

Figure 1. The results of manual checks of random samples of REF2021 articles recorded by Scopus as funded (listing a university or funder) or unfunded for six UoAs.

## 3.3 Analyses

For RQ1, the proportion of articles declaring research funding was calculated for each UoA and Main Panel.

For RQ2 and RQ3, the average quality of the articles from each funder was calculated and compared to the average quality of unfunded research. The Grade Point Average (GPA) was used for this, which is the arithmetic mean of the quality scores. Although widely used in the UK in rankings of institutions, the GPA is a convenience and not theoretically informed because there is no reason to believe that a 4* article is four times as good as a 1* article. Nevertheless, it at least gives a straightforward and easily understandable indicator of average quality for funded journal articles. The 30 largest funders (including unfunded and unknown funder) were reported. The choice of 30 is relatively arbitrary. The GPA for small funders with a few articles would be imprecise estimates of the funder's average research quality, and 30 is a common statistical choice for the minimum size to identify a pattern. This calculation ignores funders not reported by the Scopus API, which particularly effects articles with multiple funders.

We used ordinal regression to answer RQ4, with research quality as the dependant variable and research funding (binary) and the log of the number of authors as independent variables. A similar approach has previously been used with citations as a proxy for research quality as the dependant variable (Ebadi & Schiffauerova, 2016). We excluded 23 articles with no authors listed from the regressions. We ran a separate regression for each UoA and Main Panel and for all the data. Ordinal regression only assumes that the four quality scores are in ascending order but does not assume that they are equidistant, so it is better than types of regression requiring a scalar output (Gutiérrez et al., 2015). By including both authors and funding as independent variables, the regression output can show whether one of the two is redundant in any area. The log of the number of authors was used instead of the number of authors because the relationship between author numbers and log-transformed citation counts is approximately logarithmic (Thelwall & Maflahi, 2020), and the shape is similar for the relationship between REF scores and author numbers (confidential research finding).

For RQ5, we calculated a field normalised citation score for every REF2021 article to allow fair comparisons between articles from different fields. For this, we first log normalised each citation count with $\ln(1+x)$ to reduce skewing in the dataset caused by very highly cited articles (Thelwall & Fairclough, 2017). Then, we calculated the average of the logged citations for each of the 330 Scopus narrow fields and each year 2014-18 (i.e., 5x330 averages) and divided each article's logged citation count by the average for its narrow field and year. Articles in multiple fields were instead divided by the average of the relevant field averages. This gives a Normalised Log-transformed Citation Score (NLCS) (Thelwall, 2017) for each journal article. These can be compared between fields and years since, by design, a score of 1 always reflects an average number of citations for the field and year of an article. Averaging the NLCS of all articles associated with a funder gives the funder's Mean NLCS (MNLCS), which is a measure of the normalised average citation rate for the journal articles it funded. Again, an MNLCS above 1 always reflects a funder that tends to fund articles that are more cited than average for their fields and years. The most recent two years were excluded from the calculation to give a citation window of at least two years, to reduce the influence of short citation windows. Although a three-year citation window is better (Wang, 2013), it would reduce the amount of data and the log transformation in the NLCS formula reduces the statistical variability caused by shorter time windows.

# 4  Results

## 4.1  *RQ1: Prevalence of research funding*

Just under two thirds (63%) of journal articles submitted to REF2021 had funding information recorded by the Scopus API, with substantial disciplinary differences (Table 1, Figure 2). This figure excludes funded journal articles where the funder was not recorded by the author, the journal did not allow a funding declaration, or a technical issue prevented Scopus from finding the declaration (see Figure 1). This also includes research that was internally funded, whether nominally (part of the scholar's job to research) or more substantially, such as with money for equipment or research assistants. Some universities (e.g., University of Wolverhampton, not in the data set) now require scholars to record their employer as the funder within the internal research information management system for articles not externally funded, and this may encourage them to report the same within their articles.

Funding is the norm for Main Panel A (80%) and B (76%), but half as prevalent in Main Panel C (40%) and D (32%). The difference is presumably due to the need for equipment and large teams in the health, life, and physical sciences, whereas expensive or perishable equipment is probably rarer in the social sciences, arts, and humanities, except for long-term purchases (e.g., musical instruments). Moreover, there may be more social sciences, arts and humanities topics that can be researched in small teams or alone. The three UoAs with the highest proportions of funded papers are Biological Sciences (91%), Physics (91%), Clinical Medicine (88%), and Chemistry (87%). All these subjects have subfields that do not need expensive equipment: theoretical physics, theoretical chemistry, biostatistics (related to medicine), and systems biology. Thus, the result may reflect "cheaper" specialties being rare in the UK or globally.

Table 1. Number of articles, unfunded articles, and university funded articles. Number of funders per UoA, main panel or all.

| Set | UoA or Main Panel | Articles | Unfunded articles | Uni funded articles | Funders | Funders with 5+ articles |
|---|---|---|---|---|---|---|
| A | 1: Clinical Medicine | 9916 | 1173 | 262 | 844 | 134 |
| A | 2: Public Health, Health Services & Primary Care | 3890 | 745 | 130 | 401 | 75 |
| A | 3: Allied Health Prof., Dentistry, Nursing & Pharm | 9675 | 2885 | 701 | 1045 | 166 |
| A | 4: Psychology, Psychiatry & Neuroscience | 8173 | 2271 | 390 | 672 | 111 |
| A | 5: Biological Sciences | 6376 | 576 | 244 | 592 | 86 |
| A | 6: Agriculture, Food & Veterinary Sciences | 3147 | 653 | 178 | 446 | 52 |
| B | 7: Earth Systems & Environmental Sciences | 3724 | 541 | 198 | 456 | 51 |
| B | 8: Chemistry | 3274 | 426 | 172 | 321 | 43 |
| B | 9: Physics | 4499 | 396 | 98 | 272 | 43 |
| B | 10: Mathematical Sciences | 5111 | 1402 | 245 | 424 | 50 |
| B | 11: Computer Science & Informatics | 4646 | 1565 | 250 | 438 | 57 |
| B | 12: Engineering | 16335 | 4395 | 1000 | 1095 | 195 |
| C | 13: Architecture, Built Environment & Planning | 2582 | 1225 | 207 | 334 | 37 |
| C | 14: Geography & Environmental Studies | 3439 | 947 | 266 | 467 | 52 |
| C | 15: Archaeology | 545 | 156 | 50 | 123 | 14 |
| C | 16: Economics & Econometrics | 1762 | 856 | 132 | 216 | 28 |
| C | 17: Business & Management Studies | 11853 | 8210 | 759 | 810 | 117 |
| C | 18: Law | 1864 | 1442 | 80 | 153 | 13 |
| C | 19: Politics & International Studies | 2502 | 1610 | 150 | 245 | 21 |
| C | 20: Social Work & Social Policy | 3295 | 1779 | 209 | 334 | 34 |
| C | 21: Sociology | 1498 | 727 | 85 | 175 | 21 |
| C | 22: Anthropology & Development Studies | 977 | 443 | 84 | 160 | 18 |
| C | 23: Education | 3337 | 2028 | 186 | 308 | 37 |
| C | 24: Sport & Exercise Sciences, Leisure & Tourism | 2812 | 1753 | 205 | 327 | 37 |
| D | 25: Area Studies | 524 | 329 | 31 | 86 | 7 |
| D | 26: Modern Languages & Linguistics | 962 | 583 | 49 | 111 | 10 |
| D | 27: English Language and Literature | 768 | 592 | 30 | 69 | 7 |
| D | 28: History | 1082 | 769 | 47 | 91 | 9 |
| D | 29: Classics | 111 | 82 | 6 | 17 | 2 |
| D | 30: Philosophy | 806 | 559 | 49 | 75 | 9 |
| D | 31: Theology & Religious Studies | 185 | 155 | 3 | 15 | 2 |
| D | 32: Art and Design: History, Practice & Theory | 1117 | 693 | 65 | 145 | 12 |

| D | 33: Music, Drama, Dance, Perform. Arts, Film | 544 | 380 | 19 | 52 | 7 |
| D | 34: Comm. Cultural & Media Stud. Lib & Info Man | 1020 | 724 | 49 | 90 | 11 |
| A | Main Panel A | 39248 | 7925 | 1804 | 4107 | 438 |
| B | Main Panel B | 36614 | 8610 | 1925 | 1825 | 335 |
| C | Main Panel C | 35634 | 20819 | 2361 | 1858 | 328 |
| D | Main Panel D | 7071 | 4842 | 345 | 456 | 52 |
| All | All | 113877 | 41649 | 6297 | 4107 | 882 |

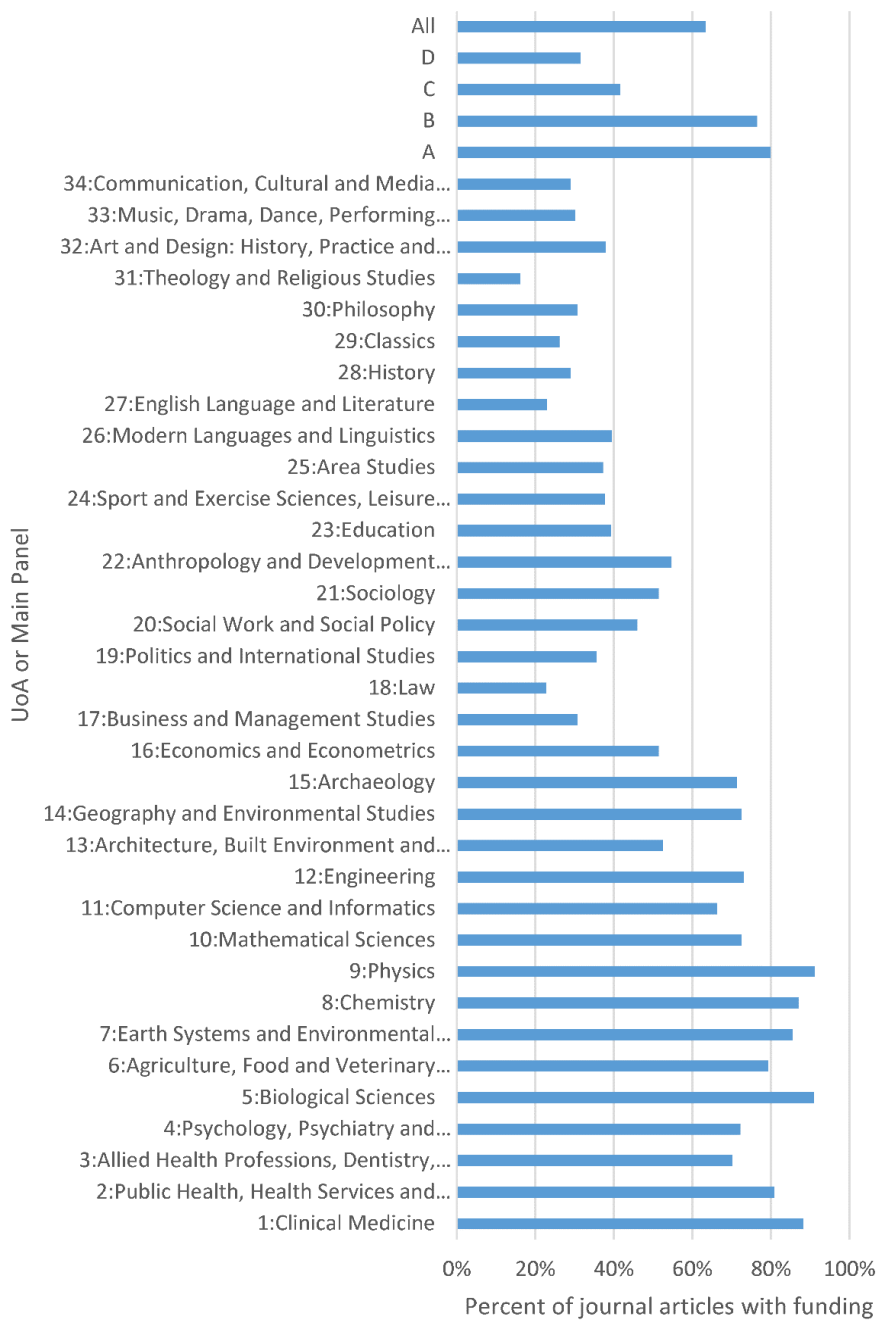

Figure 2. The percentage of UK REF2021 journal articles with a declared source of funding in Scopus.

## 4.2 RQ2: Is funded research higher quality for all major research funders?

The Grade Point Average (GPA) of the REF2021 scores of journal articles tends to be higher than the unfunded article GPA for most large research funders in Main Panels A-D (Figure 3, 4, 5, 6). In the few cases where the funded GPA is lower than the unfunded GPA, the confidence interval for the former almost always includes the latter. The sole minor exception is the European Commission funding in Main Panel C (Figure 5). Nevertheless, this exception could be a side effect of the large number of tests (29x4), and with a Bonferroni correction the difference between European Commission funded research and unfunded research in Main Panel C is not statistically significant. Thus, at the Main Panel level, the results are consistent with research funding being an advantage for all major funders, albeit marginal in some cases.

For Main Panel A (Figure 3), all research funder GPAs are above the unfunded GPA and none of the research funder confidence intervals contain the unfunded GPA. Thus, funding from a major funder is an advantage in Main Panel A. The same is broadly true for Main Panel B (Figure 3) except that four of the funder GPA confidence intervals contain the unfunded score.

The pattern is mixed for Main Panel C (Figure 5), perhaps because of smaller sample sizes giving less accurate mean estimates and wider confidence intervals. Although there are three funders with GPAs below the unfunded GPA, there are many funders with GPAs substantially above it and with narrow confidence intervals. Thus, there is still a general trend for major funder money to be advantageous in Main Panel C. For Main Panel D, most funders have a GPA above the unfunded GPA, and a few have substantially higher GPAs with narrow confidence intervals, suggesting that major funder money is also an advantage here.

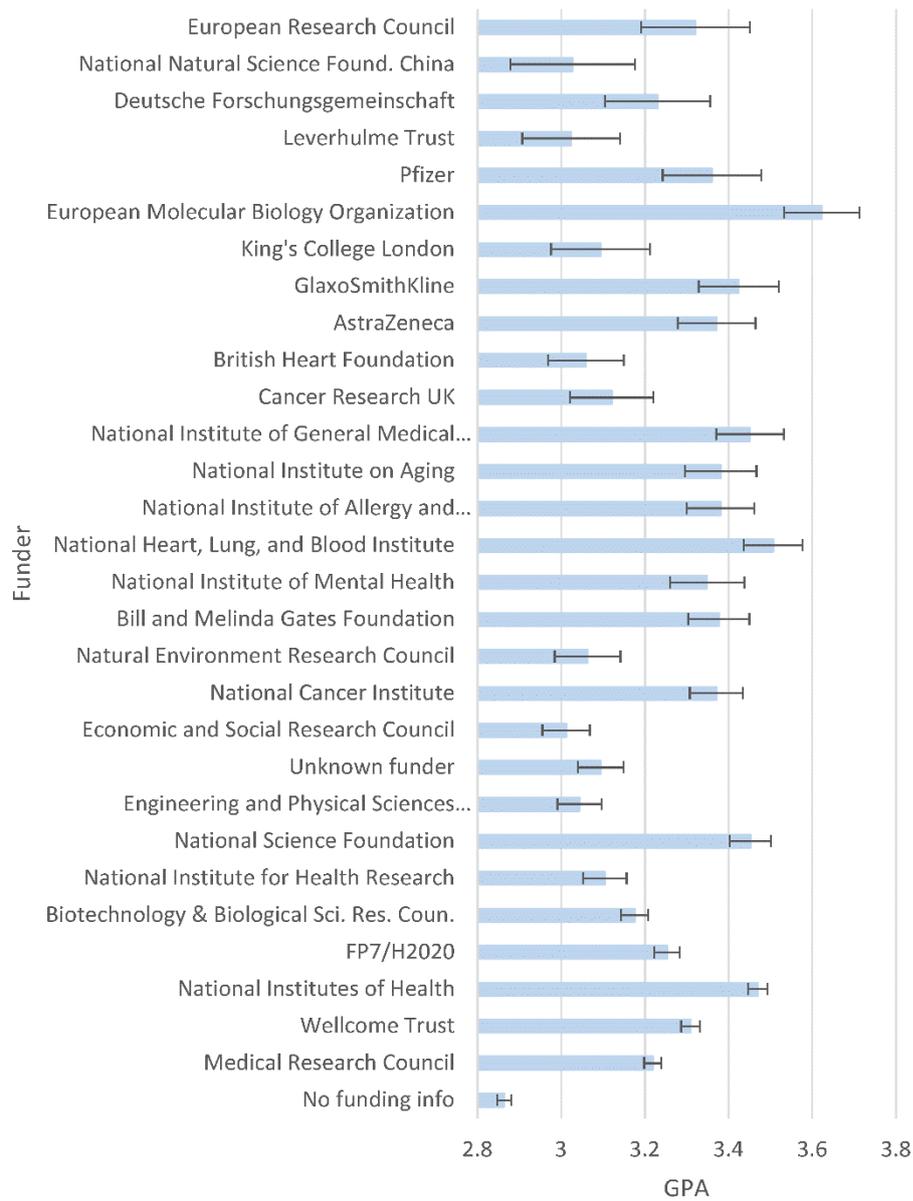

Figure 3. The average quality score of REF2021 journal articles by research funder for **Main Panel A** for 30 research funders with the most articles. Error bars indicate 95% confidence intervals.

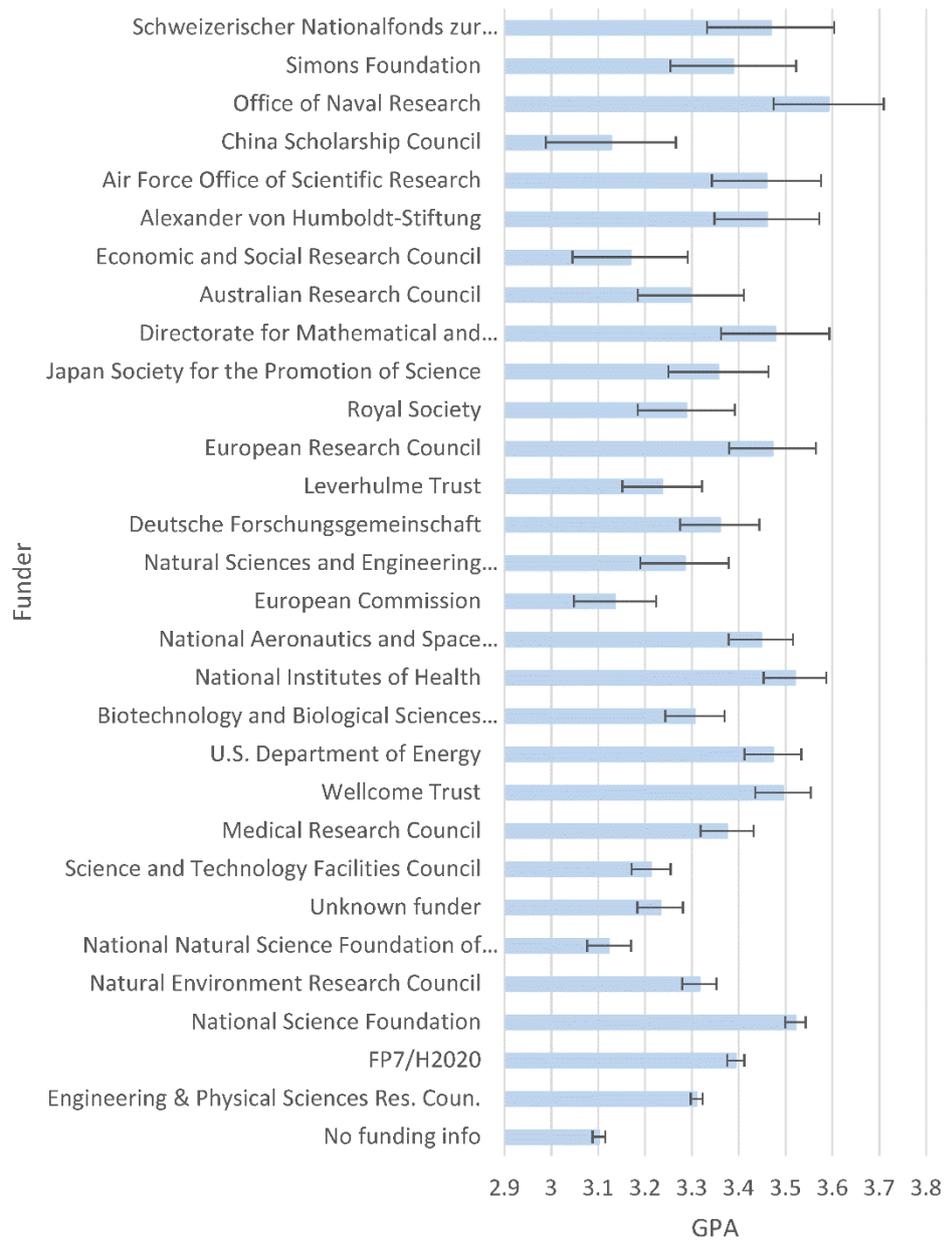

Figure 4. The average quality score of REF2021 journal articles by research funder for **Main Panel B** for 30 research funders with the most articles. Error bars indicate 95% confidence intervals.

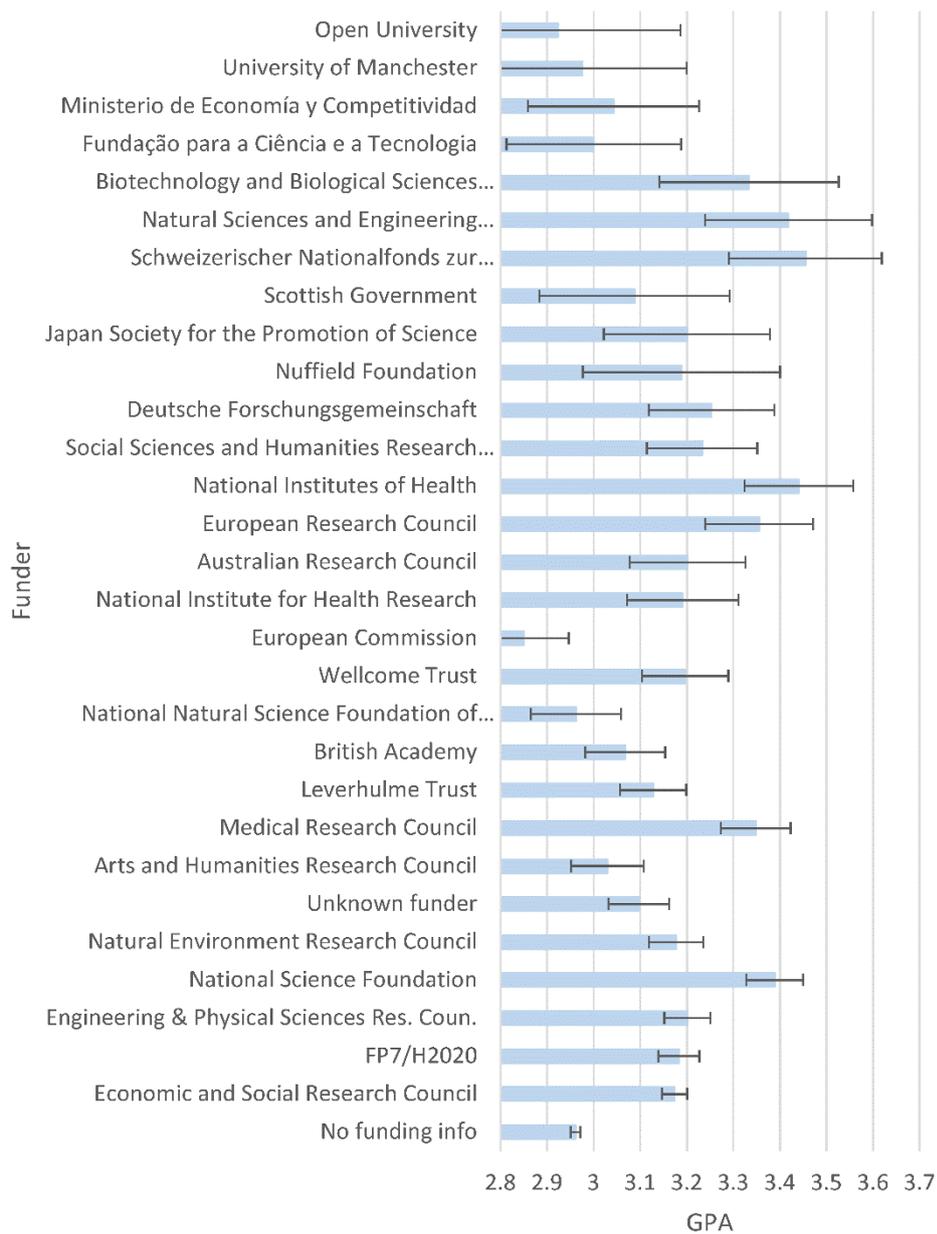

Figure 5. The average quality score of REF2021 journal articles by research funder for **Main Panel C** for 30 research funders with the most articles. Error bars indicate 95% confidence intervals.

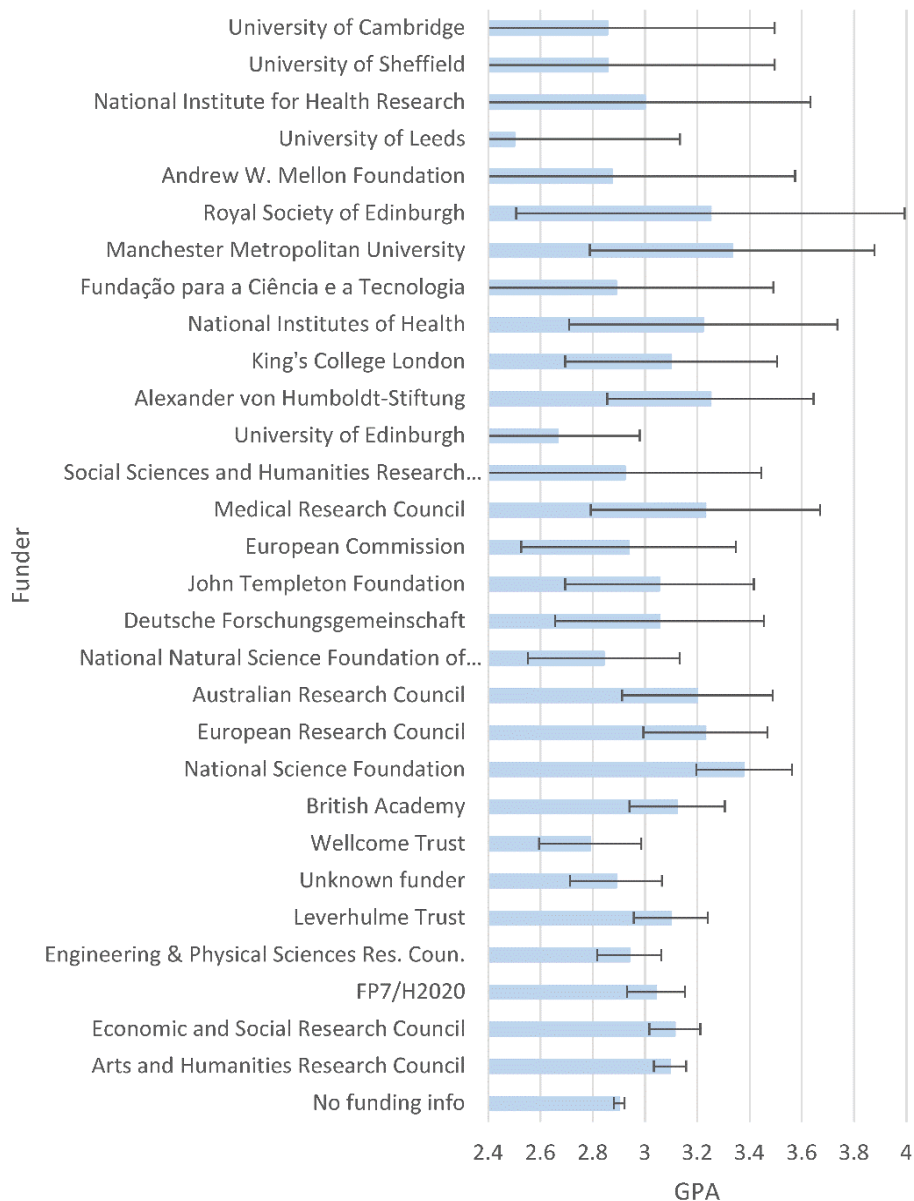

Figure 6. The average quality score of REF2021 journal articles by research funder for **Main Panel D** for 30 research funders with the most articles. Error bars indicate 95% confidence intervals.

Major funders also tend to support higher quality research when the data is aggregated to the level of individual UoAs, although there are some exceptions. For Clinical Medicine (UoA 1, Figure 7), Engineering and Physical Sciences Research Council (EPSRC) funded research surprisingly generated lower quality scores than unfunded research. The reason for this may be that UoA 1 assessors did not value research with substantial inputs from non-medical fields in the context of their UoA. There is no similar problem for UoAs 2 (Figure 8) and 3 (Figure 9).

     The presence of pharmaceutical companies as funders for health and medical research is clear in UoAs 1-3 (Figures 7, 8, 9). The research that they fund tends to have a substantially higher GPA than unfunded research, suggesting that the commercial income enhances rather than compromises academic quality.

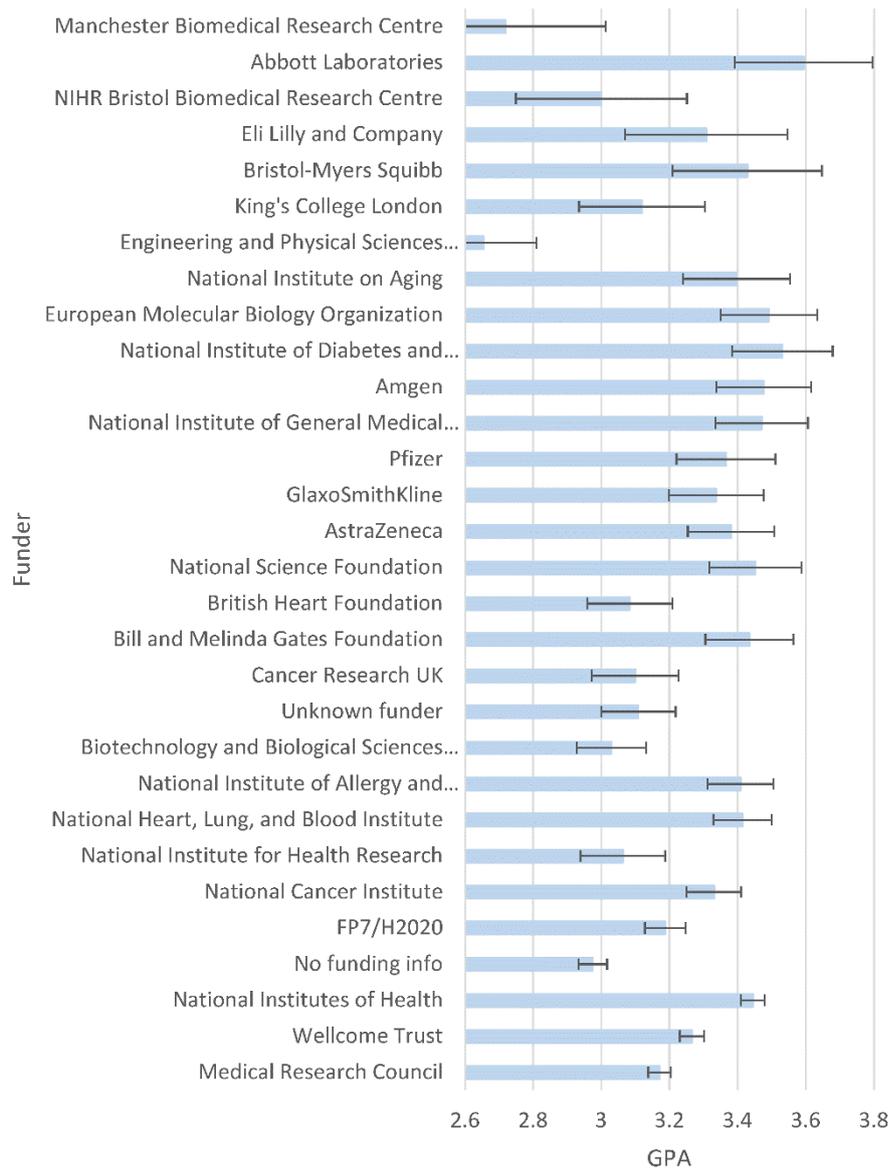

Figure 7. The average quality score of REF2021 journal articles by research funder for **UoA 1 Clinical Medicine** for 30 research funders with the most articles. Error bars indicate 95% confidence intervals.

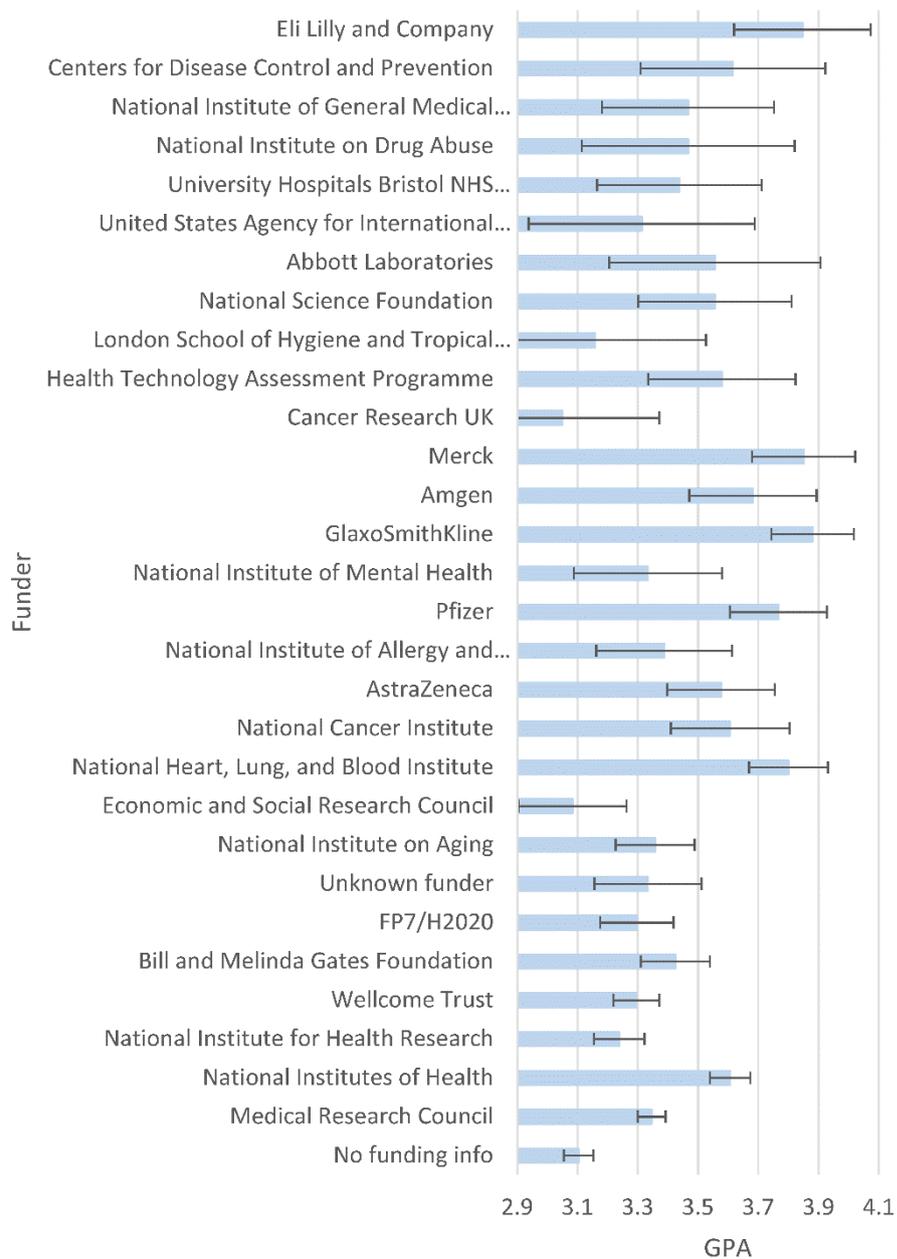

Figure 8. The average quality score of REF2021 journal articles by research funder for **UoA 2 Public Health, Health Services and Primary Care** for 30 research funders with the most articles. Error bars indicate 95% confidence intervals.

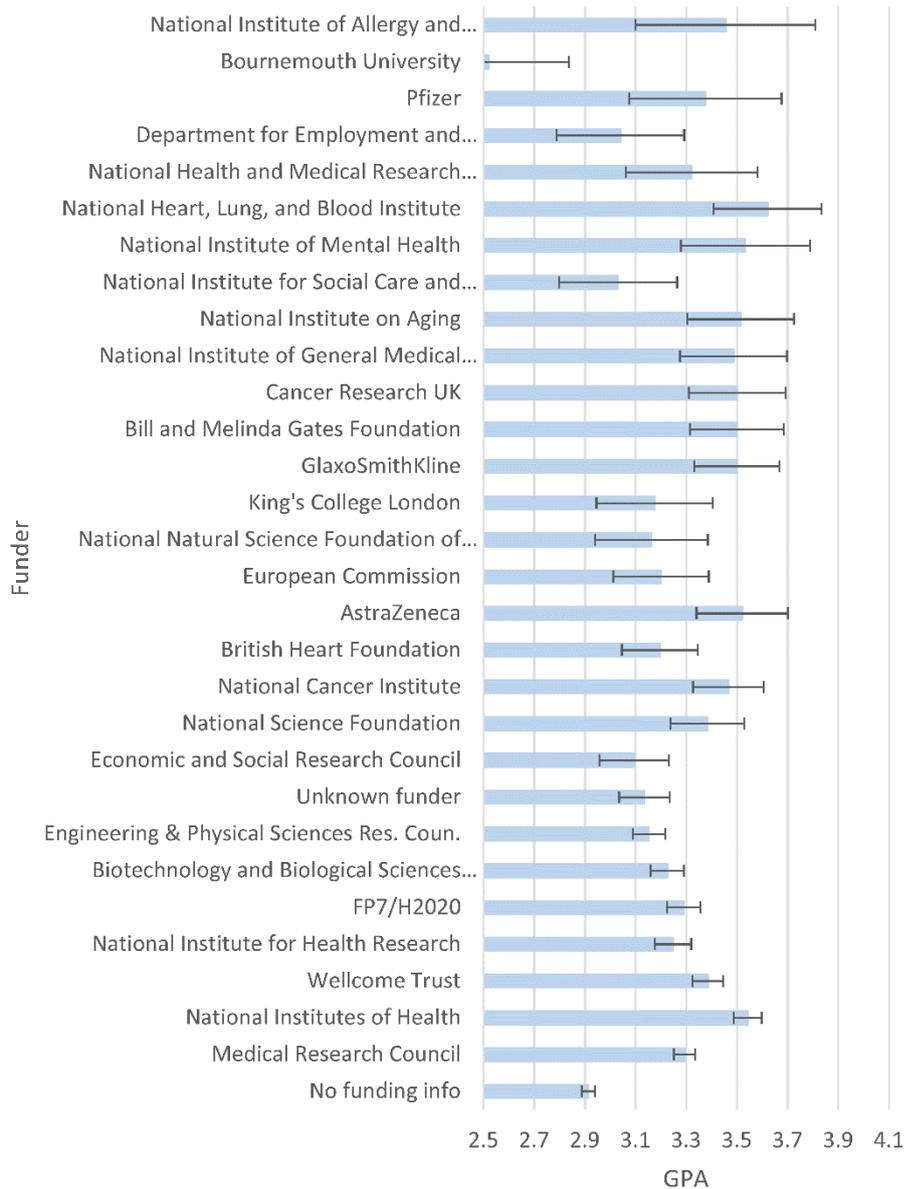

Figure 9. The average quality score of REF2021 journal articles by research funder for **UoA 3 Allied Health Prof., Dentistry, Nursing & Pharmacy** for 30 research funders with the most articles. Error bars indicate 95% confidence intervals.

The second largest UoA, Engineering (Figure 10) illustrates the general advantage of major research funders for quality in this field. Although most of the funders are governmental research funding bodies, military funding clearly produces above average quality research.

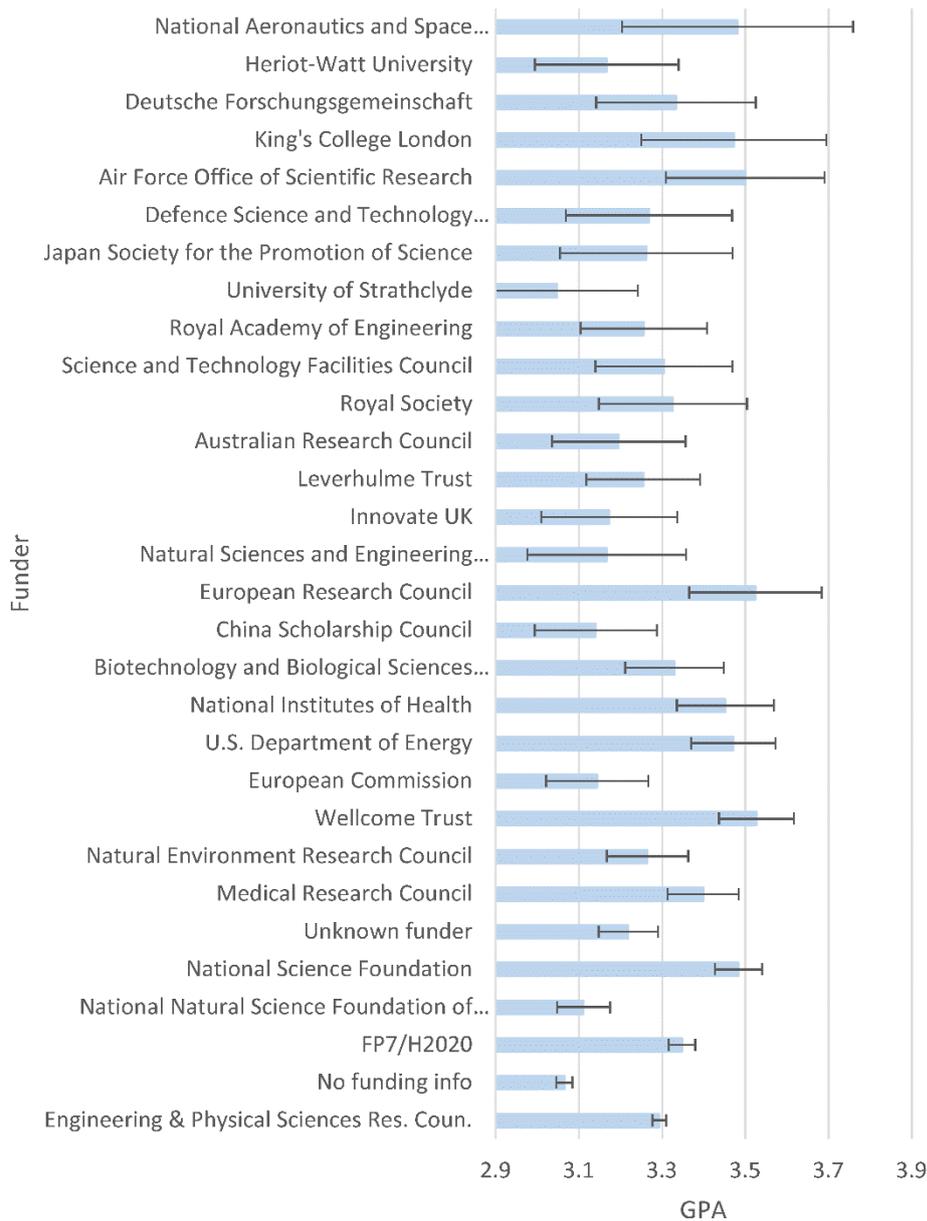

Figure 10. The average quality score of REF2021 journal articles by research funder for **UoA 12 Engineering** for 30 research funders with the most articles. Error bars indicate 95% confidence intervals.

Funding seems to be a marginal advantage for largest UoA, Business and Management, since 8 of the top 28 funders have a below average GPA (Figure 11). Moreover, the core funder, the Economic and Social Research Council (ESRC), confers the relatively minor advantage of a 0.1 higher average GPA. The European Research Council was (pre-Brexit) particularly effective at funding high quality research, but this is a logical side-effect of its strategy of selecting "top researchers" through very competitive grants.

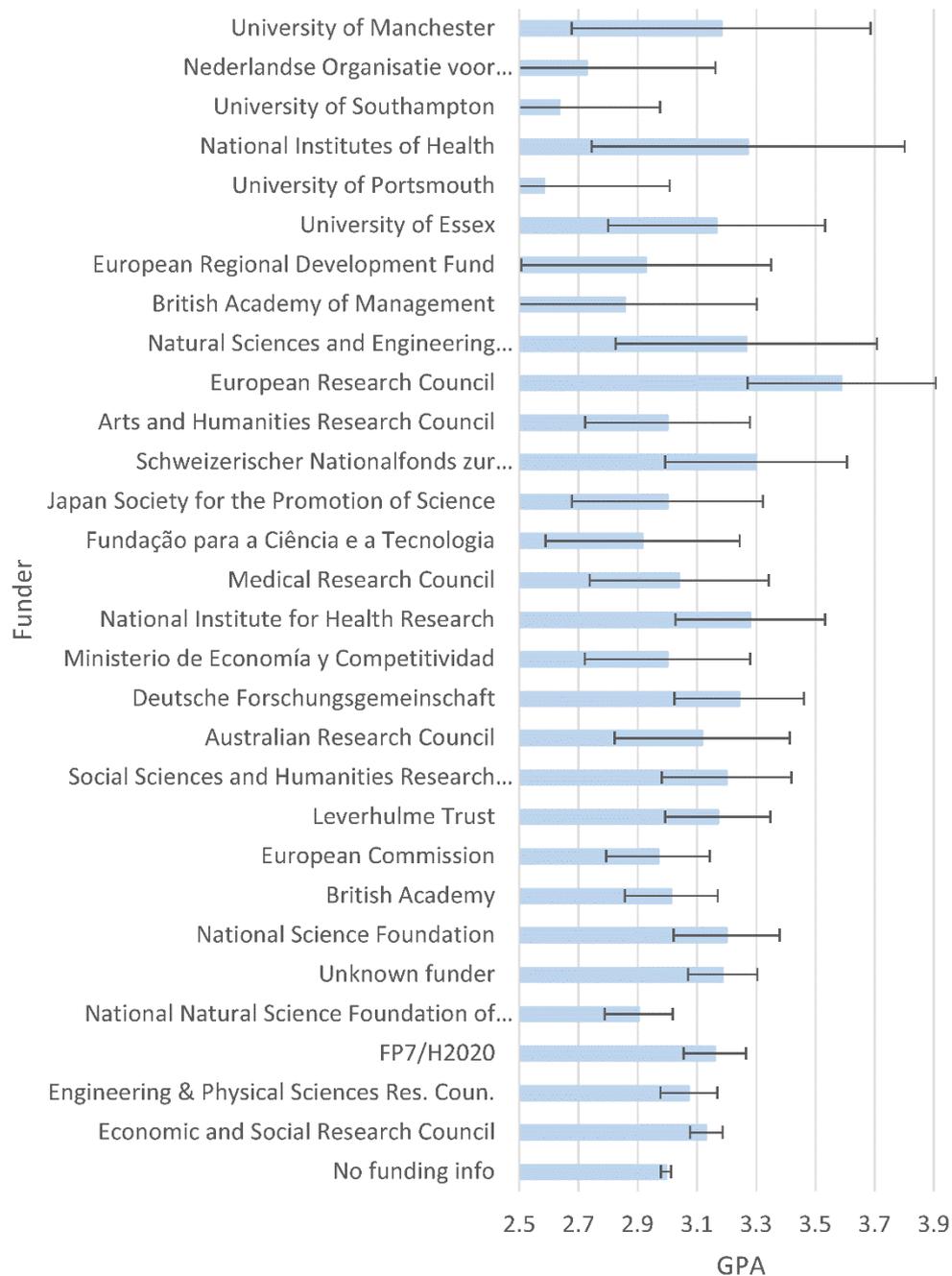

Figure 11. The average quality score of REF2021 journal articles by research funder for **UoA 17 Business and Management Studies** for 30 research funders with the most articles. Error bars indicate 95% confidence intervals.

### 4.3 RQ3: Do research funders support different quality research?

As the graphs above illustrate, there are statistically significant differences in the average quality of research supported by different funders. For example, in UoA 1 (Figure 7) the average GPA of the main three research funders is different, with their confidence intervals not overlapping. In particular, the National Institutes of Health (NIH) funded particularly high-quality research, followed by the Wellcome Trust (UK charity) and the Medical Research Council (MRC), all of which have large budgets and general funding remits. The NIH advantages may be that its research funded with UK partners would usually be international, since it is based in the USA, and its funding is backed by the greater financial resources of the USA.

### 4.4 RQ4: Does team size moderate the effect of funding on research quality?

Previous research has shown that articles with more authors tend to be more cited, and funding seems to attract large team sizes, so it is possible that the advantage of funding is sometimes primarily in bringing together a large team. In our data, for all UoAs and Main Panels, funder GPA correlates positively with the average (geometric mean) number of authors on papers associated with the funder (Figure 12, GPA vs. authors). The correlations tend to be strong in Main Panels A and B.

For UoAs with at least 30 funders associated with at least 5 papers each, the weakest correlation between GPA and authors is for Business and Management Studies (0.06). Thus, for Business and Management Studies research funders there is almost no relationship between funded collaboration size and average research quality. This may be due to relatively little variation in GPA between funders and typically small research teams (average 3 authors per paper for all major funders, varying between 1.7 and 4.3). In contrast, the highest correlation is for Agriculture, Food and Veterinary Sciences (0.84), partly due to medical funders (MRC, NIH, Wellcome) and the Gordon and Betty Moore Foundation supporting large team research with high GPAs.

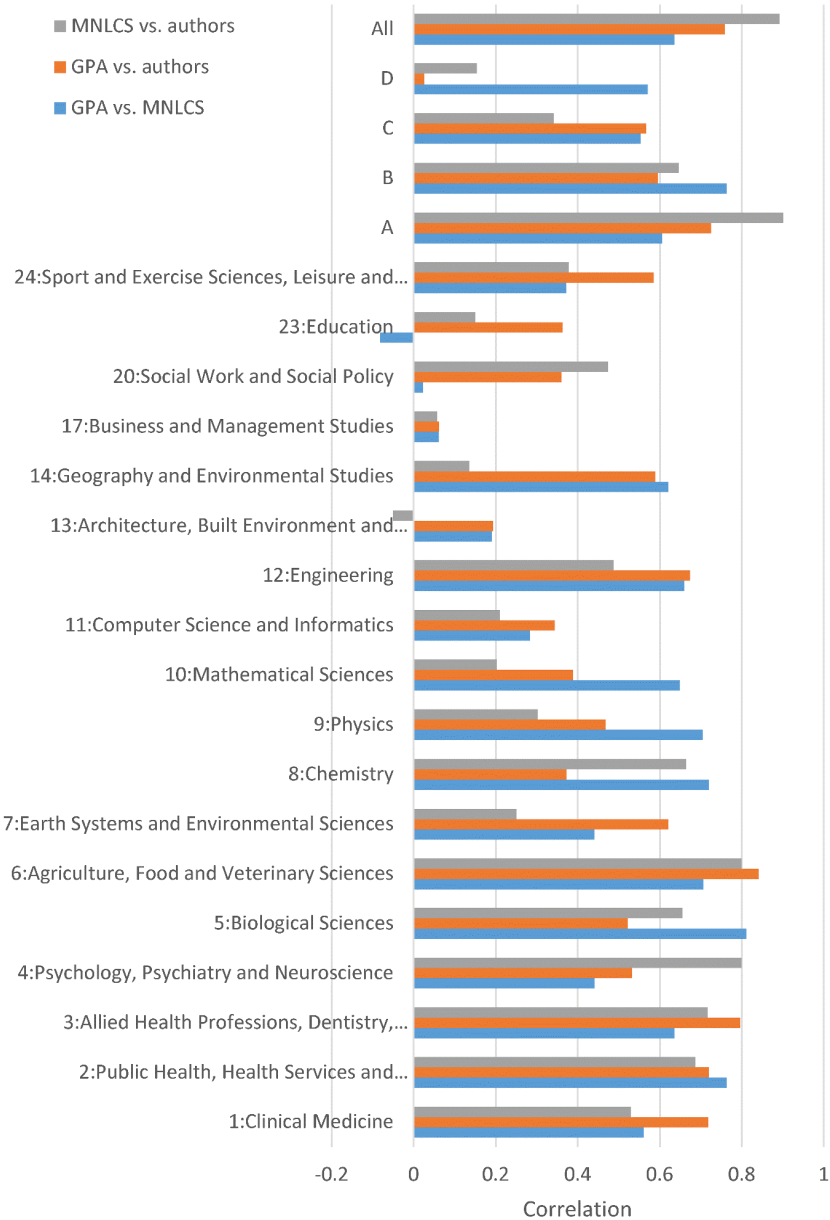

Figure 12. Pearson correlations between funder MNLCS, GPA and geometric mean authors by UoA or Main Panel. MNLCS correlations only cover research published 2014-18. UoAs are included only when they have at least 30 funders associated with at least 5 papers each.

Ordinal regressions for each UoA, Main Panel and overall (39 regressions in total) allow the effects of funding and author numbers to be analysed separately. As a conservative step (see Figure 1), university-funded research was classed as unfunded, so the focus is on external funding for research. Because of the incompleteness of the funding data, the results will tend to underestimate any differences that exist. In the regressions, an exponentiated coefficient of 1 indicates that the independent variable (logged number of authors or external funding) has no effect on the odds ratios for quality scores (1, 2, 3, or 4). Values greater than 1 indicate that the variable increases the odds ratio for a higher quality score and values less than 1 the opposite. Every increase of 1 in the exponentiated regression coefficient for funding increases the odds ratio for higher quality research by 1 for funded research compared to unfunded research. Similarly, every increase of 1 in the exponentiated regression coefficient for logged authors increases the odds ratio for higher quality research by 1 for research with *e*=2.718 times more authors.

The results show that, when considered independently from the number of authors, funding associates with improved odds of higher quality research in 33 out of 34 UoAs and all 4 Main Panels (Figure 13). The confidence intervals exclude the null value 1 for 30 out of 39 regressions. On this basis, it is plausible that, after factoring out the number of authors, funding always associates with an improved chance of higher quality journal articles. Nevertheless, whilst funding has the most substantial association with quality in Main Panels A and B, its association is marginal in some UoAs from Main Panels C and D. Thus, overall, there is evidence that funding has weak or moderate value, above generating larger team research, in associating with higher quality research in the social sciences, arts and humanities, but there is strong evidence that it has a considerable value in medicine, life and physical sciences, and engineering.

Although of less importance here, increased author numbers usually, but not always associates with increased odds of higher quality journal articles, even after factoring out research funding. The exceptions are mainly in the arts and humanities.

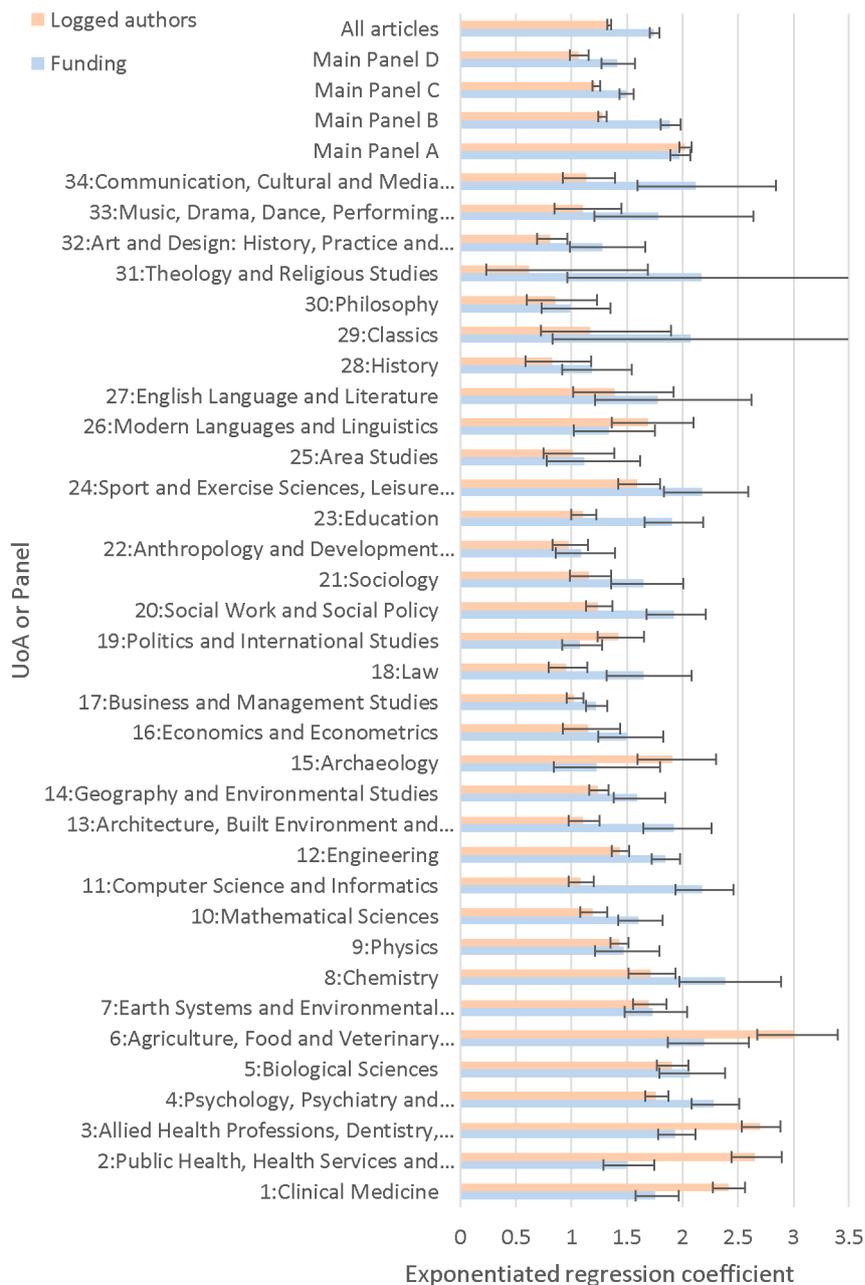

Figure 13. Exponentiated ordinal regression coefficients for Quality score against external funding (binary) and the logged number of authors for REF2021 articles 2014-20. Error bars show 95% confidence intervals. University-funded research and research without declared funding is classified as unfunded.

### 4.5 RQ5: Are average citation counts effective proxies for average quality for externally funded research?

Research funders often have their own evaluation teams to assess the effectiveness of their grants. For this, the main quantitative evidence is likely to be citation data, although they may also have project grades from end-of-grant reviewers in some cases. If they make comparisons against similar funders to make like-for-like comparisons, then the only quantitative data that they would have for both would be citation counts. Thus, it is useful to check whether the average citation impact of funders is an effective proxy for the average quality of the research that they fund.

Correlations between funder citation rates (MNLCS) and average quality (GPA) are strong (>0.5) in all Main Panels (Figure 12), suggesting that citation impact is a reasonable proxy for research quality overall. The correlations also tend to be moderate or strong in the UoAs of Main Panels A and B (Figure 12), but are variable in the UoAs of Main Panel C. In particular, the correlations are close to 0 (positive or negative) in UoAs 17 (Business and Management Studies), 20 (Social Work and Social Policy) and 23 (Education) and weak (0.2) in UoA 13 (Architecture, Built Environment and Planning). Thus, citation rates are inappropriate proxies for funder quality in these areas. By extension, and due to a lack of evidence, it seems that citation rates should not be used as proxies for funder research quality throughout the social sciences, arts, and humanities, except for Geography and Environmental Studies.

# 5 Discussion

The results are limited to journal articles from the UK, and to the best 1-5 journal articles written by academics, so are not representative of typical UK research (especially books). They only consider the funder reported by the Scopus API, ignoring any that Scopus could not find and all funders except one in the case of multiple-funded articles. This is a substantial limitation, as discussed in the evaluation at the end of the Methods section. In particular, the extent of funding is underestimated in the data here. This does not invalidate the findings because funded research is still more likely to be recorded as such in the API (Figure 1), so the funded and unfunded groups are statistically distinct. This limitation nevertheless indicates that differences found between funded and unfunded research are larger than shown in the data (because the unfunded subsets are "polluted" with funded articles).

The findings ignore the value of each grant, whether the funding was partial, what the money was spent on, how many publications were produced from it, and whether journal articles were the primary outcome of the project or a side-effect. They also ignore disciplinary differences in the need to record funding sources, with biomedical fields apparently most affected due to a need to register any potential conflicts of interest. They also ignore the purpose of the funding, which may not be to conduct high quality research but to develop a technology for industry, to train a PhD student, to develop a junior postdoc, to build research networks, or to support researcher mobility. Funding here is tied to publications, although a team may be partly funded and draw on different sources (Aagaard et al., 2021).

## 5.1 Comparison with prior work

The findings mostly have little directly comparable prior work. For RQ1, the prevalence of research funding for any country is reported apparently for the first time, albeit with partial data. The existence of disciplinary differences in funding rates is unsurprising but does not seem to have been previously investigated for all academic fields. The prevalence of funding is much higher than previously reported (Berman et al., 1995; Borkowski et al., 1992; Ernst et al., 1997; Jowkar et al., 2011; Lim et al., 2012; Shandhi et al., 2021; Stein et al., 1993), with a few exceptions (Godin, 2003), probably at least partly due to more systematic funding reporting now, and the UK sample (excluding training and publishing practitioners).

The higher quality rates for major funders (R2) is a new finding, but echoes many previous studies of individual funders that has shown funded articles or researchers to be more cited than a comparable group (unfunded articles, unsuccessful applicants, or researchers before the funding) (Álvarez-Bornstein et al., 2019; Berman et al., 1995; Gush et al., 2018; Heyard & Hottenrott, 2021; Levitt, 2011; Lewison & Dawson, 1998; Peritz, 1990;

Rigby, 2011; Roshani et al., 2021; Yan et al., 2018), and conflicts with the few studies not showing this or showing the reverse in specific fields or contexts (Jowkar et al., 2011; Muscio et al., 2017; Neufeld, 2016). The discrepancies include two fields where citations are reasonably reliable indicators of quality, Biology/Biochemistry and Environment/Ecology in Iran (Jowkar et al., 2011), so it is possible that there are international differences in the value of research funding.

The unsurprising finding that funders can support different quality research (RQ3) aligns with prior findings that research funders can support research with different average citation impacts (Thelwall et al., 2016), and that the amount of research funding influences the citation impact of the research (Muscio et al., 2017).

The finding that funded research is higher quality than unfunded research even after factoring out team size (RQ4) is not directly comparable to prior studies. It contradicts claims that the current managerial approach to research in higher education reduces the quality of research in the social sciences by restricting the autonomy of researchers (Horta & Santos, 2020), although it is not clear whether academics with more autonomy but the same amount of funding would produce better work. The evidence of fields in which average citation counts are effective proxies for average quality for externally funded research (RQ5) is also not directly comparable to prior studies.

## 5.2 Alternative causes of funded research being higher quality

The higher quality of funded research has multiple possible causes, all of which may be true to some extent. Although it seems self-evident that funding improves research, it is not always true (Jowkar et al., 2011; Muscio et al., 2017; Neufeld, 2016). There are many pathways that could explain the usually positive relationship.

**Funders select more successful researchers to fund**: Previous research, albeit with limited scope, suggests that funding councils may be good at excluding weak researchers but not good at identifying the very best, at least if citations are accepted as a proxy for research quality (Van den Besselaar et al., 2009). Assuming that the first group, together with researchers that were unable to submit funding bids, formed a majority or were substantially weaker than the other two groups, this would likely translate into a statistical association between funding and researcher quality.

**Funding improves existing research**: At the simplest level, funding may allow some researchers to conduct better versions of the research that they had already intended to pursue. For example, the funding might support a larger scale survey, newer equipment, expert collaborators, or additional supporting analyses. It seems unlikely that a project given extra funding would often become worse, for example because new equipment was bought but did not work well, or an expanded survey incorporated lower quality data collection methods in the additional areas.

**Funding changes the research carried out, replacing weaker (or no) with stronger work**. Funding might allow a study that would be impossible for the applicant(s) without external funding (Bloch et al., 2014). If the funding was for expensive equipment or other processes (e.g., large scale in-person interviews) then the work seems likely to be more original than average, assuming that few researchers in a field would have access to funding for investigations with a similar purpose. For example, perhaps an Alzheimer's researcher gets funding to run a large-scale genetic screening test and produces one of the few studies on this topic. Originality is one of the three components of research quality (Langfeldt et al., 2020), so increasing this would be enough to improve the overall quality grade for an article.

Of course, funded types of research could also sometimes tend to be weaker than unfunded research in some fields or contexts. For example, funding commonly supports PhD projects (Ates & Brechelmacher, 2013), and PhD research could be better or worse than average, depending on the field.

**Funding-led research goals are more valued**: Research projects that align with funders' strategic priorities may be highly valued if assessors accept these priorities. Although there are open call grants, some may pursue unfunded research because of the freedom to choose their own priorities (Behrens & Gray, 2001; Cheek, 2008), so strategic goals seem likely to be more common in funded research. Funding also generates an implicit hierarchy of research value, with even unfunded goals aligning with societal needs potentially being undervalued (Frickel et al., 2010).

**Funding is regarded as a good in itself**: Given high levels of competition for research funding, a funding declaration may be seen as an important achievement in itself, especially since the evaluators are mainly from a UK higher education environment in which funding is encouraged and rewarded. Conversely, in funding-rich areas, articles lacking funding may be treated with extra suspicion.

**Funding entails impact requirements**: Whilst industry funding typically has commercial value as a goal, research council grants have societal impact requirements and give resources to achieve these through dissemination activities. Thus, funded research may be more impactful through multiple pathways related to the funding sources.

# 6 Conclusions

In the UK, there are substantial disciplinary differences in the proportions of funded research and the extent to which funded research tends to be higher quality than unfunded research. Although only evaluated in a limited UK context, the results suggest, but do not prove, that there are no broad fields of research in which funding does not help academics to produce higher quality research, so no fields can afford to ignore it. The main exceptions are a few individual funders in some contexts. Moreover, since the results could be equally explained by better researchers being more successful at attracting funding or funding improving the researchers' outputs, so no cause-and-effect can be claimed. The results are not due to funded research tending to involve larger teams because the regressions showed a residual funding advantage after taking into account team size. Overall, however, since the results are at least consistent with research funding adding value apparently universally across disciplines, avoiding grants seems like a risk for all researchers, unless they have good reasons to believe that their research is an exception.

This study does not take into account productivity so the results cannot be used for a cost-benefit analysis of funding. More detailed research that considers the amount of funding available for each study and the role of the funding (e.g., improving existing research, allowing expensive studies) would be needed to make a reasonable cost-benefit analysis to give useful information about the disciplinary differences in the effectiveness of funding, but this seems unlikely to be possible with current public data.

A secondary finding is that citations are not always effective proxies for average funder quality, especially in the social sciences, arts, and humanities. Funders and studies that use citations as proxies for quality to assess the impact of funding should only do so for the fields identified above where appropriately field normalised citation counts correlate at least moderately with quality.

**AUTHOR CONTRIBUTIONS**
Mike Thelwall: Methodology, Writing–original draft, Writing–review & editing.
Kayvan Kousha: Writing–review & editing.
Mahshid Abdoli: Writing–review & editing.
Emma Stuart: Writing–review & editing.
Meiko Makita: Writing–review & editing.
Cristina I Font-Julián: Writing–review & editing.
Paul Wilson: Writing–review & editing.
Jonathan Levitt: Methodology, Writing–review & editing.

**COMPETING INTERESTS**
The authors have no competing interests to declare.

**FUNDING INFORMATION**
This study was funded by Research England, Scottish Funding Council, Higher Education Funding Council for Wales, and Department for the Economy, Northern Ireland as part of the Future Research Assessment Programme (https://www.jisc.ac.uk/future-research-assessment-programme). The content is solely the responsibility of the authors and does not necessarily represent the official views of the funders.

**DATA AVAILABILITY**
The raw data was deleted before submission to follow UKRI policy for REF2021. More data information is available in an associated report (http://cybermetrics.wlv.ac.uk/TechnologyAssistedResearchAssessment.html).